\documentstyle[12pt,svoss2,psfig]{book}

\begin{document}

\title{High-Mass Microblazars Associated with Variable Gamma-Ray Sources?}

\author{Marina Kaufman Bernad\'o, Gustavo E. Romero}

\affil{Instituto Argentino de Radioastronom\'{\i}a, C.C.5, (1894) Villa Elisa, Buenos Aires, Argentina}

\author{I. Felix Mirabel}

\affil{CEA/DMS/DAPNIA/Service d'Astrophysique, Centre d'Etudes de Saclay, 91191 Gif-sur-Yvette, France}%\\Instituto de Astronom\'{\i}a y f\'{\i}sica del Espacio/CONICET, C.C. 67, Suc 28, Buenos Airegyus, Argentina

\begin{abstract}
A model to explain variable gamma-ray sources with high-mass microblazars is described.
Inverse Compton interactions between the jet of a microquasar and the UV-photon field of the stellar companion can produce high-energy gamma-rays. The interaction with photons from the disk and the corona is also taken into account. The long-term variability presented by some sources is related to the precession of the jet, which is induced by the gravitational effect of the companion upon the accretion disk. Observations of repeated soft gamma-ray outbursts from Cyg X-1 can be consistently interpreted under this model. 
\end{abstract}


%\vspace{1cm}

\section{Scenario}

The third EGRET catalog (Hartmann et al. 1999) contained, when published, $\sim$170 gamma-ray sources that could not be clearly identified with potential counterparts at lower energies. 
These unidentified sources can be separated into two broad groups. The first one, located at low Galactic latitudes ($\left| b \right|<10^\circ$), is well-correlated with the spiral arms and the Galactic plane. It is formed by bright population I sources (Romero 2001). The other group, at middle or high latitudes, is in turn composed by three different populations: an isotropic component, a group of weak and hard sources probably related to the Gould Belt, 
and a halo component of soft sources (Grenier 2001). The identification of these sources is one of the most important challenges of contemporary high energy astrophysics.

\begin{figure}[t]
\centerline{%\hbox{
\hspace{5cm}\hbox{\psfig{figure=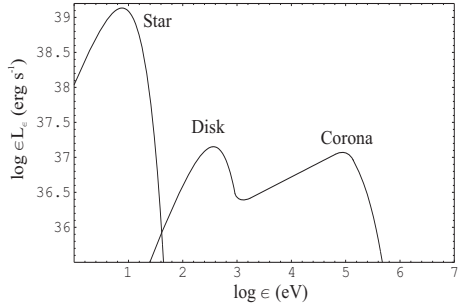,height=4 cm,width=6 cm}}}
\caption{\textit{Left}: Sketch of the proposed situation (see jpeg extra file). \textit{Right}: Prototype example of the spectrum from the external photon fields to which the jet is exposed. 
\label{figure1.eps}}
\end{figure}

In this paper we outline a model that can explain many of the characteristics of some low-latitude gamma-ray sources on the basis of presumed {\sl microblazar} activity in distant Galactic microquasars. A microblazar phase occurs when the jet of a microquasar (basically an X-ray binary with relativistic plasma ejections) points nearly toward the observer. We shall argue that the gravitational effects of the high-mass companion star can induce a precession of the relativistic beam in most of these systems. When the original viewing angle is not too large, the object can go through phases of extremely enhanced activity (the {\sl microblazar} phase) resulting in a strong gamma-ray source in the observer's frame. The gamma-ray emission is produced by inverse Compton (IC) interactions between relativistic leptons in the jet and seed photons from the star, the accretion disk, and the hot corona or ADAF region around the compact object. 

\section{Model}

As we mentioned, microquasars are X-ray binary systems able to generate relativistic jets. These jets can be detected through their synchrotron emission at radio wavelengths. We can say that microquasars are Radio Emitting X-ray Binaries (REXBs). Microblazars are microquasars with jets forming a small angle with the line of sight. They should present a significant enhanced non-thermal flux due to Doppler boosting (Mirabel \& Rodr\'{\i}guez 1999). 

In what follows we shall consider high-mass microquasars. The jet, which is created close to the central accreting object, must traverse several external photon fields. Figure 1, left panel, sketches this situation, whereas the right panel shows an example of the typical spectra of these fields. First, we have a power-law X-ray isotropic field that can be modeled as a corona or an ADAF region, with a hard spectrum. The IC interactions are in this region under the Klein-Nishina regime for MeV-GeV electrons. Then, we have the field of photons coming from the cold accretion disk, represented by a black body emission with a peak at a few keVs. Interactions with this field occur under Thomson regime. Finally, we have a nearly isotropic field due to the stellar companion, typically a black body with a peak in the UV. IC interactions occur here also with Thomson cross section.

\begin{figure}[h]
\centerline{\hbox{
\psfig{figure=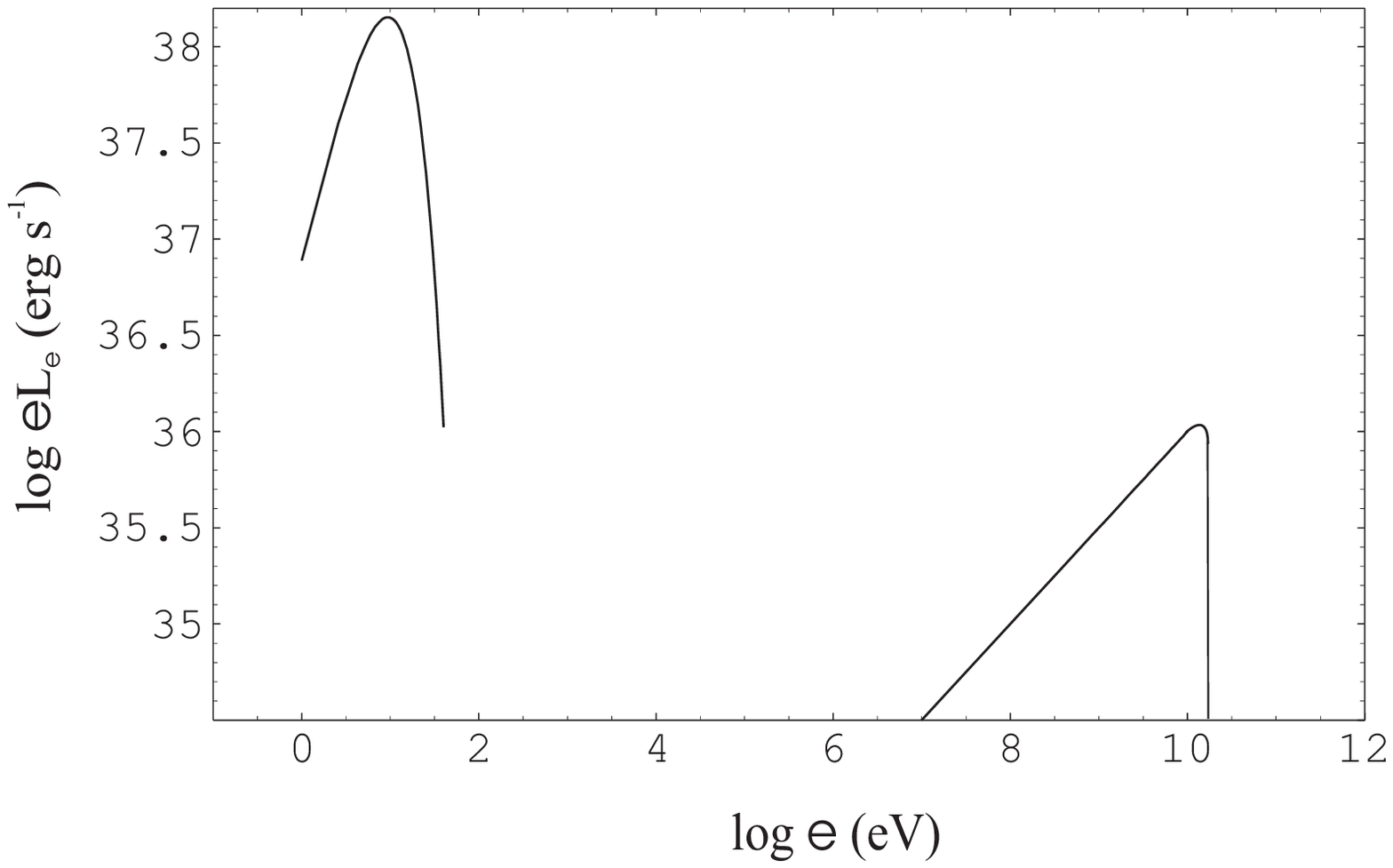,height=3.9 cm,width=6 cm}}\hspace{1cm}\hbox{\psfig{figure=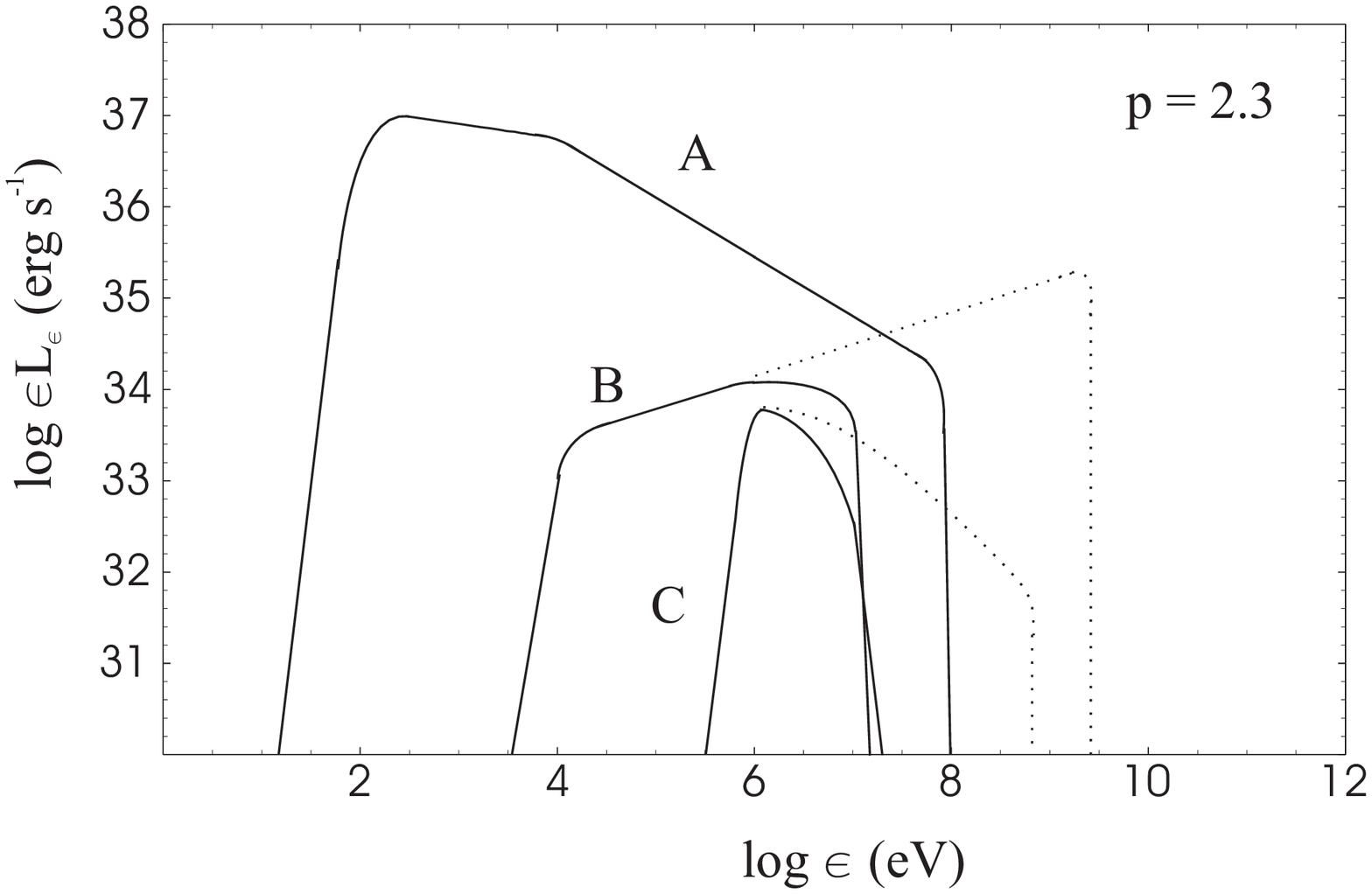,height=4.1 cm,width=6 cm}}}
\caption{\textit{Left}: Spectral high energy distribution of the scattered photons coming from the companion star. The black body spectral energy distribution of the star is also shown (left top corner) \textit{Right}: Spectral energy distribution for the up-scattering of the star (A), disk (B), and coronal (C) photons. Radiation absorbed in the local photon fields is shown in dashed lines.   
\label{figure2.eps}}
\end{figure}

For the stellar photons, the resulting IC luminosity, is given by (Geor-ganopoulos et al. 2002):

\begin{equation}
\frac{dL}{d\epsilon d\Omega}\approx D^{2+p}\frac{k V \sigma_{T} c U 2^{p-1}}{\pi\epsilon_{0}\left(1+p\right)\left(3+p\right)}\left(\frac{\epsilon}{\epsilon_{0}}\right)^{-\left(p-1\right)/2},
%\eqno{(1)}
\end{equation}
where $D = [\Gamma (1 - \beta \cos \phi ) ]^{-1}$ is the Doppler factor, $\phi$, the viewing angle, $\epsilon$, the final photon energy, $\sigma_{T}$, the Thomson cross section, $U$, the energy density of the photon field, $k$, the constant in the particle energy distribution given by $N(E)\propto E^{-p}$, and $p$, the spectral index.
 
Since the disk photons come from behind the jet, we have to add an extra factor $(1-\cos \phi)^{-(p+1)/2}$ that reduces the effects of beaming (Dermer et al. 1992). The IC scattering of the coronal photons is under the Klein-Nishina regime and then a numerical integration is needed. The result will not be a strict power-law in such a case.

In Figure 2, left panel, we concentrate on the luminosities obtained from the IC of the seed photons from the companion star. It shows the results obtained from the calculation of the spectral energy distribution for a specific model with $p=2$, bulk Lorentz factor $\Gamma = 5$, viewing angle $\phi = 10^{0}$, photon field of an O7 star (photon energy peaking at $\approx$ 10 eV), and a high energy cut-off of $\gamma_{max} = 10^{3.5}>> \gamma_{min}$. The right panel of the same figure shows the result of the up-scattering from the three different seed components: the star, the disk, and the corona or ADAF. In this case, the model calculations are for a specific source, Cyg X-1, taking $p=2.3$, $\phi = 30^{0}$, $\Gamma = 5$ and $\gamma_{max} = 10^{3}$ (see Romero et al. 2002 for a detailed treatment of this source).  Radiation absorbed in the local photon fields by pair production is shown in dashed lines. Compton losses in the different regions will modify the injected electron spectrum, which will introduce breaks in each spectral energy distribution.

\subsection{Variability}

The companion star in a high-mass microquasar not only provides a photon field for inverse Compton interactions, but also a gravitational field that can exert a torque onto the accretion disk around the compact object. The effect of this torque, in a non-coplanar system, is to induce a Newtonian precession of the disk. If the jets are coupled to the disk, as it is usually thought, then the precession will be transmitted to them. A time-parameterization of the jet's viewing angle is the needed, $\phi(t)$ (Abraham \& Romero 1999), which implies a time evolution of the boosting amplification factor of the gamma-ray emission: $D^{2+p} = [\Gamma (1 - \beta \cos \phi(t) ) ]^{-(2+p)}$. The viewing angle, $\phi$, and the opening angle of the cone define by the jet while precessing, $\theta$, can be seen in Figure 1 left panel. See Kaufman-Bernad\'o et al. (2002) for additional details. 

The time dependence of the amplification factor is shown in two different examples in Figure 3. The peak of the 
amplification, when an otherwise undetectable source can fall within the sensitivity of gamma-ray instruments, corresponds to the microblazar phase. It can be seen that a huge range of amplifications, going from a factor 10 to a factor  $6 \;10^{3}$, can be obtained depending mainly on the average viewing angle in the precession, which is of 10 or 30 degrees in the cases depicted in Figure 3.  

\begin{figure}[t]
\centerline{\hbox{
\psfig{figure=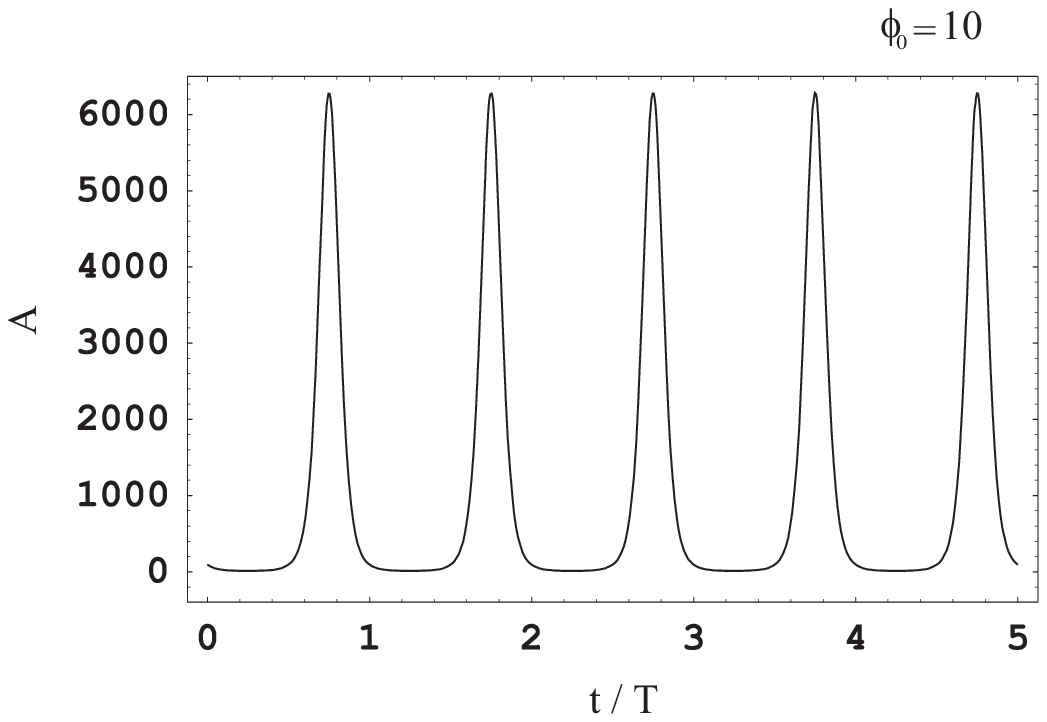,height=4.5 cm,width=6 cm}}
\hspace{1cm}\hbox{\psfig{figure=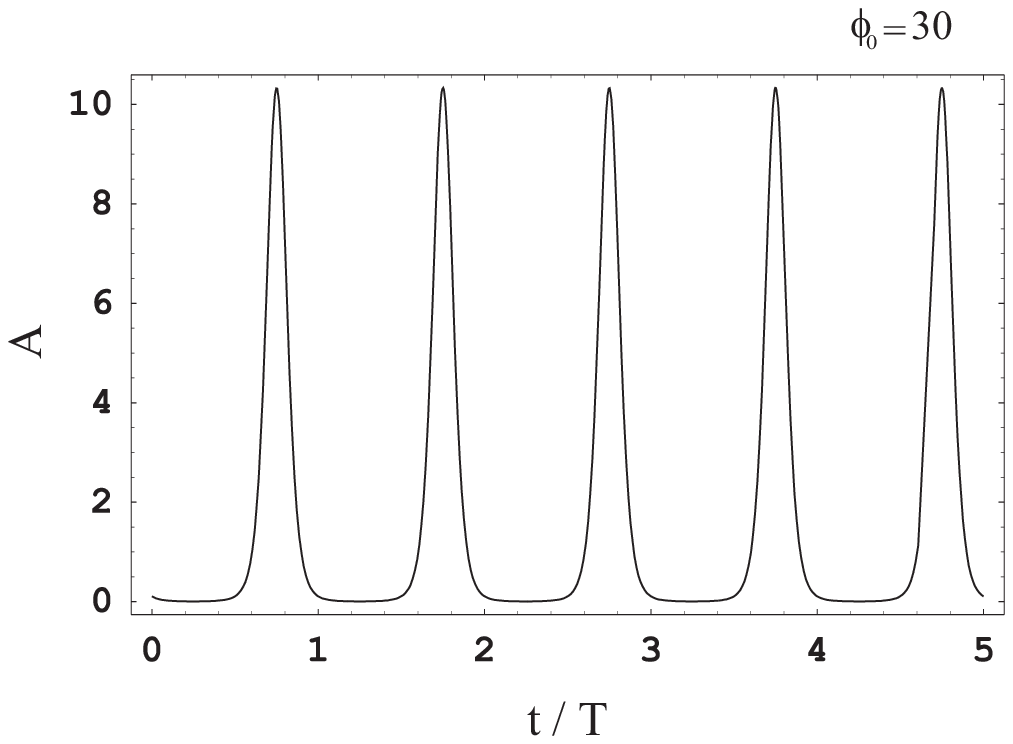,height=4.5 cm,width=6 cm}}}
\caption{Variation of the amplification factor for a continuous jet emission as a function of time in the precessing microblazar model for two different average viewing angles, $\phi_{0}$= 10 and 30 deg. Time units are normalized to the precessing period.  
\label{figure3.eps}}
\end{figure}

\section{Final comments}

It is clear from Figure 2 that in the case of a high-mass companion microblazar, the IC photons from the star dominates the observed emission. As we can see, energies in the range MeV-GeV can be reached for reasonable electron Lorentz factors of $\gamma\sim 10^3$. Nevertheless, energies of a few MeV corresponding to soft-gamma emission or hard X-rays, can also be obtained with this model as it is shown here in the specific case of Cyg X-1 (see also Georganopoulos et al. 2002), or recently suggested by Zhang et al. (2002) for the new unidentified MeV source detected by COMPTEL. The calculated luminosities are easily in the expected range $10^{35}$-$10^{36}$ erg/seg required for EGRET sources in the Galactic plane. The resulting photon energy distribution, and the luminosity inferred by the observer, will strongly depend on the assumed spectrum for the relativistic electrons of the jet and on the viewing angle $\phi$ which will determine the degree of "microblazar" of the studied microquasar.

Most of the gamma-rays produced within the coronal region will be absorbed by pair creation. The annihilation of these pairs would produce a broad, blue-shifted feature in the MeV spectrum. Because of the precession of the jet, the Doppler factor will change periodically with time, and hence the position of the annihilation peak should oscillate in energy in the lab frame around a mean value. Already launched this year, INTEGRAL satellite may be able to detect such features. On the other hand, GLAST MeV-GeV satellite, due to 2005 or 2006, will enhance the statistics on unidentified Galactic sources making possible more accurate population studies that might shed additional light on the number and distribution of microquasars in the Galaxy.
                                                                                       
\acknowledgments

High-energy astrophysics with GER is supported by Funadaci\'on Antorchas, with additional contributions from CONICET (PIP 0430/98) and ANPCT (PICT No. 03-04881). This work benefited from the Argentinian-French ECOS-SUD cooperation agreement. MK-B and GER thank the kind hospitality of the Service d'Astrophysique, Centre d'Etudes de Saclay. MK-B is grateful to the organizing committee of SuperVOSS II for financial help to participate in the meeting.

\end{document}